\documentclass[preprint2]{aastex}
\usepackage{emulateapj5}
\usepackage{onecolfloat5,epsfig}
\def\be{\begin{equation}}
\def\ee{\end{equation}}
\def\bea{\begin{eqnarray}}
\def\eea{\end{eqnarray}}

\def\cmm2{{\,\rm cm^{-2}}}
\def\cm2{{\,{\rm cm}^2}}
\def\cmm3{{\,{\rm cm}^{-3}}}
\def\gcmm3{{\,{\rm g\,cm^{-3}}}}

\def\HO{{100h\,{\rm km\,sec^{-1}\,Mpc^{-1}}}}

\def\fun#1#2{\lower3.6pt\vbox{\baselineskip0pt\lineskip.9pt
  \ialign{$\mathsurround=0pt#1\hfil##\hfil$\crcr#2\crcr\sim\crcr}}}
\def\C{{\cal C}}
\def\P{{\cal P}}



\def\hmpc{{\, {\rm h}^{-1}~\rm Mpc}}

\def\msun{{\,M_\odot}}

\def\deg{^\circ}
\def\'{^{\prime}}

\def\p3m{P$^3$M}

\def\ga{\mathrel{\mathpalette\fun >}}
\def\fun#1#2{\lower3.6pt\vbox{\baselineskip0pt\lineskip.9pt
  \ialign{$\mathsurround=0pt#1\hfil##\hfil$\crcr#2\crcr\sim\crcr}}}

\font\BF=cmmib10
\def\gam{\hat{\gamma}}

\def\x{{\hbox{\BF x}}}
\def\r{{\hbox{\BF r}}}

\lefthead{HUTERER ET AL.}
\righthead{}

\begin{document}
\twocolumn[
\submitted{Submitted to ApJ}
\title{The Angular Power Spectrum of EDSGC Galaxies} 
\author{Dragan Huterer}
\affil{Department of Physics,
University of Chicago, Chicago, Illinois 60637, USA,
email: dhuterer@sealion.uchicago.edu}
\vspace{0.5cm}
\author{Lloyd Knox\altaffilmark{1}}
\affil{Department of Astronomy and Astrophysics,
University of Chicago, Chicago, Illinois 60637, USA,
email: knox@flight.uchicago.edu}
\vspace{0.3cm}
\and
\vspace{-0.15cm}
\author{Robert C. Nichol}
\affil{Department of Physics, Carnegie Mellon University,
Pittsburgh, Pennsylvania 15213, USA,
email:  nichol@cmu.edu}

\begin{abstract}
  We determine the angular power spectrum $C_l$ of the Edinburgh/Durham
  Southern Galaxy Catalog (EDSGC) and use this statistic to constrain 
  cosmological parameters.  Our methods for determining $C_l$ and the 
  parameters that affect it are based on those developed for the 
  analysis of cosmic microwave background maps.  We expect them to 
  be useful for future surveys.
  Assuming flat cold dark matter models with a cosmological constant 
  (constrained by {\it COBE}/DMR and local cluster abundances) and a 
  scale--independent bias $b$, we find acceptable fits to the EDSGC angular 
  power spectrum with $1.11 < b < 2.35$ and $0.2 < \Omega_m < 0.55$ at 95\%
  confidence.  These results are not significantly affected by 
  the ``integral constraint'' or extinction by interstellar dust, 
  but may be by our assumption of Gaussianity.
\end{abstract}

]

\altaffiltext{1}{Currently at Institut d'Astrophysique de Paris}

\section{Introduction}

Over the next decade, the quantity and quality of galaxy survey
data will improve greatly because of a variety of new survey
projects underway including the Sloan Digital Sky Survey (SDSS;
see York et al.\ 2000).  However, most of the galaxies in such
surveys will {\it not} have spectroscopically--determined
redshifts; therefore, the study of their angular correlations
will be highly profitable for our understanding of the
large--scale structure of the Universe.

The primary purpose of this paper is to consider an analysis
approach which is likely to be useful for deriving cosmological
constraints from these larger surveys.  In particular, we use
methods that have become standard in the analysis of cosmic
microwave background (CMB) anisotropy maps, such as those from
{\it Boomerang} (de Bernardis et al.\ 2000; Lange et al.\ 2000)
and {\it Maxima}-I (Hanany et al.\ 2000, Balbi et al.\ 2000).

Estimation of the two--point angular correlation function
$w(\theta)$ from galaxy surveys without redshfit information has
a long history.  Early work (Peebles \& Hauser 1974; Groth \&
Peebles 1977) found the angular correlation function to vary as
$w(\theta)=\theta^{1-\gamma}$ with $\gamma= 1.77$ and a break at
scales larger than $\sim 9\hmpc$. The advent of automated
surveys, such as the APM galaxy survey (Maddox et.\ al.\ 1990) and
Edinburgh/Durham Southern Galaxy Catalogue (EDSGC; Collins,
Nichol \& Lumsden 1992) enabled a much more accurate
determination of $w(\theta)$, as each survey contained angular
positions for over a million galaxies.

One way to compare the measured angular correlation function with
theoretical predictions is to invert $w(\theta)$ to obtain the
three--dimensional power spectrum $P(k)$. This requires
inverting Limber's equation (Limber 1953).  Baugh \& Efstathiou
(1993, 1994) and Gazta\~{n}aga \& Baugh (1998) used Lucy's
algorithm (Lucy 1974) to do the inversion, while Dodelson \&
Gazta\~{n}aga (2000) used a Bayesian prior constraining the
smoothness of the power spectrum.  Eisenstein \& Zaldarriaga
(2000) used a technique based on singular-value decomposition to
get $P(k)$ from $w(\theta)$.  They point out that, once the
correlations in the inverted power spectra are included, the
uncertainties on cosmological parameters from the APM are
significantly weakened.

Our analysis is a three-step process, similar to what is done with 
CMB data sets (Tegmark 1997; Bond, Jaffe \& Knox 1998; 
Bond, Jaffe \& Knox 2000).  
The first step is the construction of a pixelized map of
galaxy counts, together with its noise properties.  The second step is
the determination of the angular power spectrum $C_l$ of the map
using likelihood analysis, together with window functions and
a covariance matrix.  In the final step, we compare our
observationally--determined $C_l$ to the $C_l$ predicted for a
given set of parameters in order to get constraints on those
parameters.  We assume that the errors in $C_l$ are log--normally
distributed.

The angular power spectrum $C_l$ is a useful intermediate step 
on this road from galaxy catalog to parameter constraints. 
Estimates of the angular power spectrum, together with a
description of the uncertainties, can be viewed as a form of
data compression.  One has converted the $\sim\!1$ million EDSGC
galaxies (for example) into a handful of power spectrum constraints,
together with window functions and covariance matrices.  Thus
if one wishes to make other assumptions about bias and cosmological
parameters than we have done here and determine the resulting
constraints, one can do so {\it without} having to return to the 
cumbersome galaxy catalog.  

We use $C_l$ instead of its historically--preferred Legendre transform
$w(\theta)$ for several reasons.
First, the error matrix structure is much simpler:
$\langle \delta C_l \delta C_{l'} \rangle$ is band-diagonal
and becomes diagonal in the limit of full--sky coverage, whereas
$\langle \delta w(\theta) \delta w(\theta') \rangle$ is much
more complicated and does not become diagonal even in the full--sky limit.
Second, the relation between $C_l$ and the corresponding three--dimensional
statistic $P(k)$ is simpler than that between $w(\theta)$ and $P(k)$
(or its Fourier transform $\xi(r)$ (Baugh \& Efstathiou 1994)).  

We use likelihood analysis to determine $C_l$ because the likelihood
is a fundamental statistical quantity.  The likelihood is the
probability of the data given $C_l$, which by Bayes' theorem is
proportional to the probability of $C_l$ given the data.  Another
advantage of likelihood analysis is that, as explained below,
it allows for straightforward control of systematic errors (due to, e.g.,
masking) via modifications of the noise matrix.

Only on sufficiently large scales do we expect the likelihood function
to be a Gaussian that depends only on $C_l$ and not any higher-order
correlations.  We therefore restrict our analysis to $l$ values less
than some critical value.  On small scales the likelihood function
becomes much more complicated and its form harder to predict {\it a
priori}.  Mode-mode coupling due to nonlinear evolution leads to
departures of the $C_l$ covariance matrix from band-diagonal.
Therefore some of the advantages of likelihood analysis and the
angular power spectrum are lost on smaller scales where other
techniques may be superior.  The Gaussianity assumption is perhaps the
weakest point of the approach outlined here.  Below, we briefly
discuss how the analysis can be improved in this regard with future
data sets.

The EDSGC, with over a million galaxies and covering over 1000
sq.\ degrees, offers us an excellent test bed for applying our
algorithms (Nichol, Collins \& Lumsden 2000).  We convert this
catalog into a pixelized map and determine its angular power
spectrum together with window functions and covariance matrix.
As an illustrative application of the angular power spectrum, we
constrain a scale--independent bias parameter $b$ and the
cosmological constant density parameter $\Omega_\Lambda$ in a
{\it COBE}--normalized $\Lambda$CDM model with zero mean spatial
curvature.  Our constraints on the bias are improved by including
constraints on the amplitude of the power spectrum derived from
number densities of low--redshift massive clusters of galaxies
(Viana \& Liddle 1999, hereafter VL99; also see Pierpaoli, Scott
\& White 2000).  These number densities are sensitive to the
amplitude of the matter power spectrum calculated in linear
perturbation theory, near the range of length scales probed by
the EDSGC.

The angular power spectrum of the APM catalog was previously estimated
by Baugh \& Efstathiou (1994) though not via likelihood analysis.
Very recently, Efstathiou \& Moody (2000) have applied the same techniques we
do here to estimating $C_l$ for the APM survey.  Their approach
differs from ours in how they constrain cosmological parameters.
Instead of projecting the theoretical three--dimensional power spectra
$P(k)$ into angular power spectra, they transform their $C_l$ constraints
into (highly correlated) constraints on $P(k)$ and then compare to 
theoretical $P(k)$.  

We expect the analysis methods presented here to be useful for
other current and future data sets---even those with large
numbers of measured redshifts.  For example, the Sloan Digital
Sky Survey (SDSS) will spectroscopically determine the redshifts
of a million galaxies, but there will be about one hundred times as
many galaxies in the photometric data, without spectroscopic
redshifts.  One can generalize the methods presented here to
analyze sets of maps produced from galaxies in different
photometric redshift slices.

In \S \ref{sec:likelihood}, we review likelihood analysis and the
use of the quadratic estimator to iteratively find the maximum of
the likelihood function.  In \S \ref{sec:Cl} we describe our
calculation of $P(k)$ and its projection to $C_l$.  In \S
\ref{sec:extract} we show how to compare the calculated $C_l$ to
the measured $C_l$ in order to determine parameters.  In \S
\ref{sec:edsgc} we apply our methods to the EDSGC, and discuss
some possible sources of systematic error in \S \ref{sec:sys}.
This is followed by a discussion of our results in \S
\ref{sec:discussion} and a brief conclusion in \S
\ref{sec:conclusions}.  An appendix outlines the derivation of
the projection of $P(k)$ to $C_l$.

\section{The likelihood function and quadratic estimation}
\label{sec:likelihood}

The likelihood is a fundamental statistical quantity:  the probability
of the data given some theory.  According to Bayes' theorem, 
the probability of the parameters of the assumed theory is proportional
to the likelihood times any prior probability distribution we care
to give the parameters.  Thus, determining the location of the
likelihood maximum and understanding the behavior of the likelihood
function in that neighborhood (i.e., understanding the uncertainties)
is of great interest.  

Despite its fundamental importance, an exact likelihood analysis is
not always possible.  Two things can stand in our way:  insufficient
computer resources for evaluation of the likelihood function (operation 
count scales as $N_{\rm pix}^3$ and
memory use scales as $N_{\rm pix}^2$), and, even worse, 
the absence of an analytic expression for the likelihood function.

In this paper we {\it assume} that the pixelized map of galaxy
counts is a Gaussian random field---an assumption which provides
us with the analytic expression for the likelihood function.  For
models with Gaussian initial conditions (which are the only
models we consider here), we expect this to be a good
approximation on sufficiently large scales.  Since we restrict
ourselves to studying large-scale fluctuations, we can use large
pixels, thereby reducing $N_{\rm pix}$ and ensuring that the
likelihood analysis is tractable.  We also check the Gaussianity
assumption with histograms of the pixel distribution.  On the
large scales of interest here and for a given 
three-dimensional length scale, Gaussianity is a better
approximation for a galaxy count survey than for a redshift
survey, due in part to the redshift-space distortions which
affect the latter (Hivon et al.\ 1995).  The projection from 3
dimensions to 2 dimensions also tends to decrease
non-Gaussianity.

Where likelihood analysis is possible, it naturally handles
the problems of other estimators (such as edge effects).
Likelihood analysis also provides a convenient framework for
taking into account various sources of systematic error, such
as spatially varying reddening and the ``integral constraint''
discussed in \S \ref{sec:sys}.

To begin our likelihood analysis, we assume that the data are
simply the angular position of each galaxy observed---though it
is possible to generalize the following analysis and use either
magnitude information or color redshifts.  We pixelize the sky 
and count the number of galaxies in each
pixel $G_i$.  Then we calculate the fractional deviation of
that number from the ensemble average:
\be
\Delta_i \equiv {G_i-\bar G \Omega_i \over \bar G \Omega_i}
\ee
where $\bar G$ is the ensemble average number of galaxies per unit solid
angle and $\Omega_i$ is the pixel solid angle.  
We do not actually know the ensemble mean.  In practice, we approximate it
with the survey average $\tilde G$.  We discuss this
approximation in \S \ref{sec:sys} and demonstrate that it
has negligible impact on our results.

We model the fractional deviation
in each pixel from the mean as having a contribution from ``signal''
and from ``noise'', so that 
\be
\label{eqn:model}
\Delta_i = s_i+n_i.  
\ee
The covariance matrix, $C_{ij}$, for the fractional deviation
in each pixel from the mean is given by:

\be
\label{eqn:C}
C_{ij} \equiv \langle \Delta_i \Delta_j \rangle = 
S_{ij} + N_{ij}
\ee
where $S_{ij} \equiv \langle s_i s_j\rangle$ and $N_{ij} \equiv
\langle n_i n_j\rangle$ are the signal and noise covariance
matrices.  Roughly speaking, signal is part of the data that is
due to mass fluctuations along the line-of-sight (see
the appendix), and noise are those fluctuations due
to anything else.

The signal covariance matrix $S_{ij}$ depends on the parameters of
interest (the angular power spectrum $C_l$) via:
\be
S_{ij} = w(\theta_{ij}) = \sum_l {2l+1\over 4\pi} C_l P_l\left(\cos\theta_{ij}\right)
e^{-l^2\sigma_b^2}
\label{eqn:signal_matrix}
\ee
where $\theta_{ij}$ is the angular distance between pixels $i$ and $j$ 
and we have assumed a Gaussian smoothing of the pixelized 
galaxy map with fwhm $= \sqrt{8ln2} \sigma_b$.  In practice,
we do not estimate each $C_l$ individually but binned $C_l$s
with bin widths greater than $\sim \pi / \theta$ where $\theta$ is a typical
angular dimension for the survey.

The noise contribution to the fluctuations $n$ is due to the
fact that two regions of space with the same mass density can
have different number of galaxies.  We model this additional
source of fluctuations as a Gaussian random process with variance
equal to $1/\bar G$, so that
\be 
N_{ij} \equiv \langle n_i n_j \rangle 
= 1/\bar G \,\delta_{ij}.  
\ee 
 More sophisticated modeling of the noise is not necessary
because at all $l$ values of interest, the variance
in $C_l$ due to the noise is much smaller than the sample
variance.

To find the maximum of the likelihood function, we iteratively
apply the following equation:
\be
\label{eqn:quadest}
\delta \C_l = {1\over 2} F^{-1}_{ll'} {\rm Tr}\left[(\Delta \Delta^T - C)
(C^{-1} \partial C / \partial \C_{l'}C^{-1})\right],
\ee
where $F$ is the Fisher matrix given by

\be
F_{ll'} = {1 \over 2} {\rm Tr}\left(C^{-1} 
{\partial C \over \partial \C_l}
C^{-1} {\partial C \over \partial \C_{l'}}\right),
\ee
and for later convenience we are using $\C_l \equiv
l(l+1)C_l/(2\pi)$ instead of $C_l$. That is, start with an
initial guess of $\C_l$, update this to $\C_l + \delta \C_l$, and
repeat.  We have found that this iterative procedure converges to well
within the size of the error bars quite rapidly.

The small sky coverage prevents us from determining
each multipole moment individually;  thus we determine the power
spectrum in bands of $l$ instead, call them ``band powers'', and denote
them by $\C_B$ where

\be
\C_l \equiv {l(l+1)C_l \over 2\pi} = \sum_B \chi_{B(l)} \C_B
\ee
and $\chi_{B(l)}$ is unity for $l_<(B) < l < l_>(B)$ where 
$l_<(B)$ and $l_>(B)$ delimit band $B$.  

Although we view equation (\ref{eqn:quadest}) as a means of finding the
maximum of the likelihood function, one can also treat $\C_l +
\delta \C_l$ (with no iteration) as an estimator in its own right
(Tegmark 1997; Bond, Jaffe \& Knox 1998).  It is referred to as a
quadratic estimator since it is a quadratic function of the data.
One can view equation (\ref{eqn:quadest}) as a weighted sum over
$(\Delta \Delta^T -C)$, with the weights chosen to optimally
change $\C_l$ so that $C$ is closer to $\Delta \Delta^T$ in an
average sense.

Various sources of systematic error can be taken into account
by including extra terms in the modeling of the data 
(equation (\ref{eqn:model})) and working out the effect
on the data covariance matrix, $C$.  Below we see specific examples
as we take into account the ``integral constraint'' and 
pixel masking.  The reader may also wish to see the appendix of 
Bond, Jaffe \& Knox (1998), Tegmark et al.\ 1998 and 
Knox et al.\ (1998) for more general discussions.

\section{Calculation of C$_l$}\label{sec:Cl}

We need to be able to calculate $C_l$ for a given theory, in order
to compare it with $C_l$ estimated from the data.
This calculation is a three-step process.
Step one is to calculate the matter power spectrum $P(k)$ 
in linear perturbation theory.  Step two is to then use some
biasing prescription to convert this to the galaxy number count
power spectrum $P_G(k)$.  Step three is to project this $P(k)$ to $C_l$.
We further discuss these steps in the following subsections.

\subsection{The 3D matter power spectrum, $P(k)$}

We take the primordial matter power spectrum to be a power-law
with power-spectral index $n$ and amplitude $\delta_H^2$ at the
Hubble radius.  We write the matter power spectrum today $P_0(k)$
(calculated using linear perturbation theory) as a
product of the primordial spectrum and a transfer function
$T(k)$: \be \P_0(k) \equiv {k^3 P_0(k)\over 2\pi^2} = \delta_H^2
(k/H_0)^{3+n} T^2(k) \ee where $H_0=\HO$ is the Hubble parameter
today.  The transfer function, $T(k)$, goes to unity at large
scales since causality prevents microphysical processes from
altering the spectrum at large scales.  At higher $k$ it depends
on $h$, $\Omega_m h$, and $\Omega_b h^2$.  To calculate the
transfer function we use the semi--analytic approximation of
Eisenstein \& Hu (2000).  It is also available as an output from
the publically available CMBfast Boltzmann code (Seljak \&
Zaldarriaga 1996).

Our power spectrum is now parametrized by 5 parameters: $n$,
$\delta_H$, $\Omega_m h$, $h$ and $\Omega_b h^2$.  In the
following analysis, we eliminate two of these parameters by
simply fixing $h=0.7$ and $\Omega_b h^2 = 0.019$.  The dependence
of our results on variations in $h$ can be derived analytically,
which we do in \S \ref{sec:discussion}.  
Measurements of deuterium abundances in the
Ly-${\alpha}$ forest, combined with the dependence of primordial
abundances on the baryon density, lead to the
constraint $\Omega_b h^2 = 0.019 \pm 0.002$ at 95\% confidence
(Burles \& Tytler 1998; Burles, Nollett \& Turner 2000).

Of the remaining parameters, two more, $\delta_H$ and $n$,
can be fixed by insisting on agreement with both the amplitude 
of CMB anisotropy on large angular scales as measured 
by {\it COBE}/DMR, and the number density of massive clusters
at low redshifts.  The {\it COBE} constraint can be
expressed with the fitting formula

\bea
\delta_H&=&1.94 \times 10^{-5}
\Omega_m^{-0.785-0.05\ln{\Omega_m}}\times \nonumber\\
&& \exp\left[-0.95\left(n-1\right) -0.170 \left(n-1\right)^2\right],
\eea
which is valid for the flat $\Lambda$CDM models we are considering
(Bunn \& White 1995).

The cluster abundance constraint can be expressed as a constraint
on $\sigma_8$ which is the rms fluctuation of mass in spheres of
radius $r=8h^{-1}$Mpc, calculated in {\it linear} theory:

\be
\sigma_8^2 = \int {dk\over k} \left(3j_1(kr)/(kr)\right)^2 \P_0(k)
\ee
where $j_1(x) = \left(x\cos(x)-\sin(x)\right)/x^2$.
VL99 find the most likely value
of $\sigma_8$ to be $\sigma_8 = 0.56\,\Omega_m^{-0.47}$.

The reason for the choice of the scale of $8h^{-1}$Mpc is that a
sphere of this size has a mass of about $10^{15}\msun$, which is the
mass of a large galaxy cluster.  Most of the $\Omega_m$
dependence of $\sigma_8$ comes from the fact that the
pre-collapse length scale corresponding to a given mass depends
on the matter density.  Thus, in a low density Universe the
pre-collapse scale is larger, and since there is less fluctuation
power on larger scales, the $\sigma_8$ normalization has to be
higher for fixed cluster abundance.

The shift in pre-collapse length scale with changing
$\Omega_m$ is very slow, scaling as $\Omega_m^{1/3}$.  Thus,
although the parameters that govern the shape of the power
spectrum affect the normalization, their influence is
quite small.  For example, the scale shift for changing
$\Omega_m$ by a factor of 3 is $3^{1/3} = 1.44$, and over this
range an uncertainty in $n$ of 0.2 translates into an
uncertainty in power of 8\%.

Of course there are uncertainties in both the constraint from
{\it COBE} and the constraint from cluster abundances.  More
significant of the two is the uncertainty in cluster abundance
constraint.  Consequently, we extend our grid of models to cover a
range of values of $\sigma_8^c$ where $\sigma_8 = \sigma_8^c
\Omega_m^{-0.47}$.  VL99 find that the probability of $\sigma_8^c$
is log-normally distributed with a maximum at $\sigma_8^c = 0.56$
and a variance of $\ln \sigma_8^c$ of
$0.25\ln^2(1+0.20\Omega_m^{0.2 {\rm log}_{10} \Omega_m})$.  The
{\it COBE} uncertainty is only 7\%.  We ignore this source of
uncertainty and do not expect it to affect our results since such
a small departure from the nominal large-scale normalization can
be easily mimicked, over the range of scales probed by EDSGC, by
a very small change in the tilt $n$.

\begin{figure}[htbp]
\begin{center}
\psfig{file= 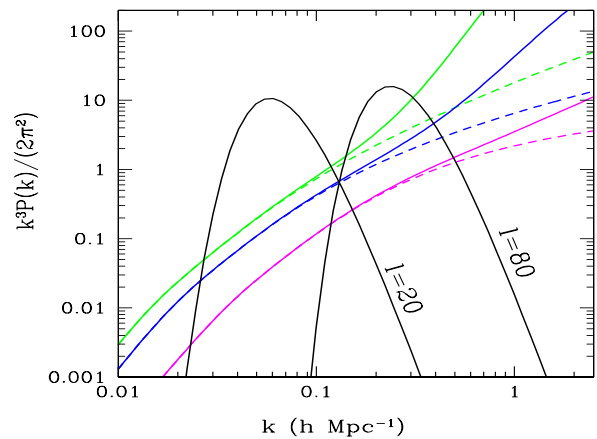, height=2.625in, width=3.4in}
\caption{Matter power spectra and $\C_l$ derivatives.  From
bottom to top at low $l$ are the COBE and cluster consistent
predictions for $\Omega_m=0.15,0.3,1$, all with $b=1$ (dashed
lines show linear theory predictions).  The $l=20$ and $l=80$
curves show $k\partial \C_l / \partial \P_k$ for for these two
multipole moments (arbitrary normalization).}
\label{fig:pk}
\end{center}
\end{figure}

In Fig.~\ref{fig:pk} we plot $\P_0(k)$ (dashed lines) for several 
models that satisfy the COBE/DMR and VL99 constraints.
Changing $\Omega_m h$ and also satisfying the $\delta_H$ and
$\sigma_8$ constraints forces $n$ to change as well.
For $\Omega_m=0.15,0.3,0.35,0.4,1$,  $n=1.55,1.00,0.91,0.84,0.47$ 
respectively.  One can understand this by considering the
simpler case of $\delta_H$ and $\sigma_8$ held constant without
$\Omega_m$ and $n$ dependence.  Then the only effect of
changing $\Omega_m h$ is to change the transfer function.  For
fixed $\delta_H$, increasing $\Omega_m h$ in this case leads
to increased power on small scales.  One therefore needs to
decrease the tilt in order to keep $\sigma_8$ unchanged. 
Now, the fact that our two amplitude constraints do depend
on $\Omega_m$ also has an effect on how $n$ changes with changing
$\Omega_m$.  However, this is a subdominant effect because these
dependences are quite similar.


\subsection{The biasing prescription}

Although biasing in general is stochastic, non--linear, redshift
and scale dependent, we adopt the simplest possible model here
in which the galaxy number density fluctuations are directly
proportional to the matter density fluctuations.  Then we can
write $b \equiv \delta_G/\delta$, where $\delta = \delta \rho
/\bar \rho$ is the matter density contrast, $\delta_G$ is the
galaxy number density contrast, and $b$ is the bias factor.

With this description $P_G(k) = b^2 P(k)$, where $P(k)$ is
the matter power spectrum.  Note that above we have only
calculated the linear theory matter power spectrum.  Non-linear
corrections are important over the EDSGC range of length
scales, and we must incorporate these effects.  We derive $P(k)$
from the linear theory power spectra 
$P_0(k)$ by use of a fitting formula (Peacock \& Dodds 1996) which
provides a good fit to the results of $n$-body calculations.
The resulting power spectra are shown by the solid lines
in Fig.~\ref{fig:pk}.

We have assumed that the galaxy number density fluctuations are
completely determined by the local density contrast.  The number
density of galaxies must also have some non--local dependence on
the density contrast.  More complicated modeling of the
relationship, or ``biasing schemes'' (e.g., Cen \& Ostriker 1992;
Mann, Peacock \& Heavens 1998, 
Dekel \& Lahav 1999) are beyond the scope of this paper.  In the applications
that follow we assume the bias to be independent of time or scale, although
our formalism allows inclusion of both of these possibilities.

From analytic theory (e.g., Seljak 2000) we expect the bias to be
scale--independent on scales that are larger than any collapsed dark matter
halos.  Numerical simulations show this to be the case as well (see Blanton et
al.\ 2000; Nayayanan, Berlind \& Weinberg 2000) on scales larger than $10
h^{-1}$Mpc. Moreover, recent observations by Miller, Nichol \& Batuski (2001)
show that a scale--independent, linear, biasing model works well when scaling
cluster \& galaxy data over the range of $200$ to $40h^{-1}$ Mpc. Our results
are determined mostly by information from these large scales.  Since we find
acceptable fits to the data using our constant bias model, we have no evidence
for a scale--dependent bias.

\subsection{The projection to 2D}

As described in the appendix, $C_l$ can be calculated from
$\P_0(k)$ and the selection function as 
\be
\label{eqn:Limber1a}
C_l = {4 \pi} \int \P_0(k) f_l(k)^2 dk/k
\ee
where
\be
\label{eqn:Limber1b}
f_l(k) \equiv {1 \over \bar G} \int dz {dr \over dz} j_l(kr)r^2\bar g(z) D(z) b T_{\rm nl}(k,z),
\ee
$r$ is the comoving distance along our past light cone, $\bar
g(z)$ is the mean comoving number density of observable galaxies,
$D(z)$ is the  growth of perturbations in linear theory relative to
$z=0$ and $T_{\rm nl}(k,z)$ is the correction factor for non-linear evolution
(Peacock \& Dodds 1996).

Equations (\ref{eqn:Limber1a}) and (\ref{eqn:Limber1b}) are 
valid for all angular scales.  It becomes
time--consuming to evaluate the Bessel function on smaller angular scales.
Although we always used equations (\ref{eqn:Limber1a}) and 
(\ref{eqn:Limber1b}), the reader should know that 
there is a much more rapid approximation which works well at $l \ga 30$:
\be
\label{eqn:Limber2}
C_l = {1 \over \bar G^2} \int dz {dr \over dz} P(k=l/r,z) [\bar g(z) D(z) b T_{\rm nl}(k=l/r,z)]^2.
\ee

In order to calculate $C_l$, we need to know $\bar g(z)$.  Since
$r^2 \bar g(z) dr/dz = d\bar G/dz$ (Baugh \& Efstathiou 1993; 1994; our
appendix) it is sufficient to know $d\bar G/dz$,
whose measurement is described in \S \ref{sec:edsgc}.

To give an idea of how $C_l$ depends on $\P(k)$ we plot 
$ k \partial C_l / \partial \P(k)$ in Fig.~\ref{fig:pk} for 
$l=20$ and $l=80$.  This quantity is the contribution to
$C_l$ from each logarithmic interval in $k$.  Note that it
is the breadth of these derivatives that explains the
correlations that appear in any attempt to reconstruct $P(k)$
from angular correlation data.  The derivatives have some dependence
on cosmology; those plotted are for the $\Omega_m=0.3$ case.

The angular power spectrum is sensitive not only to the power
spectrum today, but to the power spectrum in the past as well.
In linear theory, the evolution of the power spectrum is
separable in $k$ and $z$; one can write $P(k,z)=P(k,0)D^2(z)$
where $D(z)$ is the growth factor well-described by the fitting
formula of Carroll, Press \& Turner (1992).  We also assume that
this relation holds for the non-linear power spectra.  In truth,
non-linear evolution is more rapid at higher $k$ than at lower
$k$.  We expect our approximation to therefore be overestimates
of $C_l$, but since we do not use data that reach very far into
the non-linear regime, we do not expect this error to be
significant.

\section{Extraction of Parameters}\label{sec:extract}

To find the maximum-likelihood power spectrum, we have iteratively applied
the binned version of equation (\ref{eqn:quadest}).  Although 
equation (\ref{eqn:quadest})
is used as an iterative means of finding the maximum of the likelihood,
it is also convenient to write it as the equivalent equation 
for $\C_B$, instead of the correction $\delta \C_B$: 

\be
\C_B = {1 \over 2}\sum_{B'} F^{-1}_{BB'} 
{\rm Tr}\left[\left(\Delta \Delta^{\rm T} - N\right)C^{-1}
{\partial C \over \partial \C_{B'}}C^{-1}\right],
\ee
where the right--hand side is evaluated at the previous iteration value of
$\C_B$, $\C_B^{\rm RHS}$, and 
$\C_B = \C_B^{\rm RHS} + \delta \C_B$ is the updated power spectrum.

We have  shown how to calculate $\C_l$ from the theoretical parameters.
We now need to calculate what $\C_B$ we expect for this $\C_l$.
One can show that the expectation value for $\C_B$, given that
the data are realized from a power spectrum $\C_l$, is
\bea
\langle \C_B \rangle &=& \sum_l \sum_{B'} F^{-1}_{BB'} \sum_{l' \in B'}
F_{ll'} \C_l \nonumber \\
&=&\sum_l {W^B_l \over l} \C_l
\eea
where the Fisher matrices on the right-hand side are evaluated at
$\C_B^{\rm RHS}$, and the last line serves to define the bandpower
window function $W^B_l$. Note that the sum over $l'$ is only
from $l_<(B')$ to $l_>(B')$.  This equation reduces to equation (8)
of Knox (1999) in the limit of diagonal $F_{BB'}$.
It is this expectation value that should be
compared to the measured $\C_B$.  

As shown by Bond, Jaffe \& Knox (2000), the probability
distribution of $C_l$ is well--approximated by an offset
log-normal form.  In the sample--variance limit, which applies for
our analysis of EDSGC, this reduces to a log--normal distribution.
Therefore we take the uncertainty in each $\C_B$ to be
log--normally distributed and evaluate the following $\chi^2$:

\bea
\label{eqn:chisqEDSGC}
&& \chi^2_{\rm EDSGC}(\Omega_m,b,\sigma_8^c) = \\
&& \hspace{0.5cm} \sum_{BB'} \left(\ln{\C_B}-
\ln{\C_B^t}\right) 
\C_B F_{BB'} \C_{B'} \left(\ln{\C_{B'}}-\ln{\C_{B'}^t}\right) \nonumber \\ 
&& \C_B^t \equiv
\sum_l {W^B_l \over l} \C_l(\Omega_m,b,\sigma_8^c)
\eea

\noindent where $\sigma_8 = \sigma_8^c \,\Omega_m^{-0.47}$.  

Our total $\chi^2 = \chi^2_{\rm EDSGC} + \chi^2_{\rm VL}$ 
includes the contribution from the cluster
abundance constraint which is also log-normal:
\be
\label{eqn:chisqVL}
\chi^2_{\rm VL} = \left(\ln{\sigma_8^c} - \ln{0.56}\right)^2/\sigma^2
\ee

\noindent where $\sigma={1\over 2}\ln{\left(1+
0.32\,\Omega_m^{0.24{\rm log}_{10}{\Omega_m}}\right)}$ (Viana and
Liddle 1996, hereafter VL96).
Note that here and throughout we have adopted the more
conservative uncertainty in VL96,  as opposed to the VL99
uncertainty.

\newpage
\section{Application to the EDSGC}\label{sec:edsgc}

The Edinburgh/Durham Southern Galaxy Catalogue (EDSGC) is a sample of nearly
1.5 million galaxies covering over $1000\ {\rm deg^2}$ centered on the South
Galactic Pole. The reader is referred to Nichol, Collins \& Lumsden (2000)
for a full description of the construction of this galaxy catalogue as well as
a review of the science derived from this survey. For the EDSGC data, the
reader is referred to www.edsgc.org.

\begin{figure}[htbp]
\begin{center}
\psfig{file=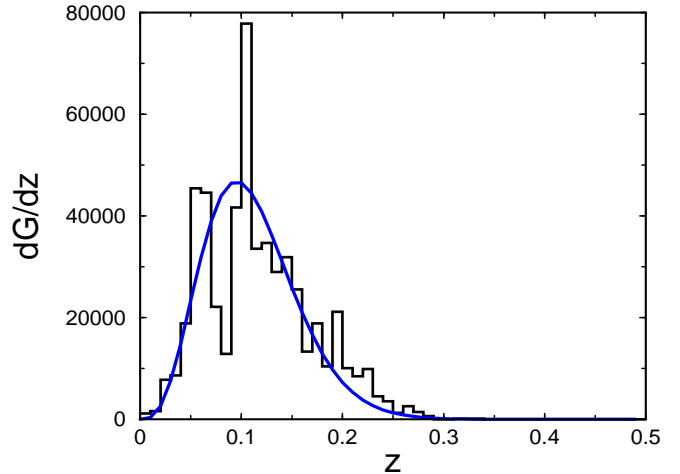 , height=2.625in, width=3.4in}
\caption{The selection function for the EDSGC; i.e., the
mean number of galaxies per steradian per redshift interval. 
}
\label{fig:selection}
\end{center}
\end{figure}

For the analysis discussed in this paper, we consider only the
contiguous region of the EDSGC defined in Nichol, Collins \&
Lumsden (2000) and Collins, Nichol \& Lumsden (1992) (right
ascensions $23< \alpha <3$ hours, through zero hours, and
declinations $-42\deg< \delta <-23\deg$).  We also restrict the
analysis to the magnitude range $10 < b_J<19.4$.  The faint end
of this range is nearly one magnitude brighter than the
completeness limit of the EDSGC (see Nichol, Collins \& Lumsden
2000) but corresponds to the limiting magnitude of the ESO Slice
Project (ESP) of Vettolani et al.\ (1998) which was originally
based on the EDSGC. The ESP survey is 85\% complete to this
limiting magnitude ($b_j=19.4$) and consists of 3342 galaxies
with redshift determination. This allows us to compute the
selection function of the whole EDSGC survey which is shown in
Fig.~\ref{fig:selection}. The data shown in this figure has been
corrected for the 15\% incompleteness in galaxies brighter than
$b_j=19.4$ with no measured redshifts as well as the mean stellar
contamination of 12\% found by Zucca et al.\ (1997) in the
EDSGC. These corrections are not strong functions of magnitude;
therefore, we apply them as constant values across the whole
magnitude range of the survey.

As mentioned above, we need to correct our power spectrum
estimates for stellar contamination in the EDSGC map.  If the
stars are uncorrelated (which we assume) then their presence will
suppress the fluctuation power as we now explain.  Let $T_i$ be
the total count in pixel $i$, consisting of galaxies and stars:
$T_i=G_i+S_i$ (for simplicity, we consider equal-area
pixels). Let $\alpha=0.12$ be the fraction of the total that are
stars, so that $\bar G=(1-\alpha)\bar T$.  Then, defining
$\Delta_i^G=(G_i-\bar G)/{\bar G}$ and $\Delta_i^S=(S_i-\bar
S)/{\bar S}$, we have
\bea 
\Delta_i &\equiv& {T_i-\bar T \over \bar T} \\
&=& (1-\alpha)\Delta_i^G + \alpha\Delta_i^S.
\eea
$\Delta_i^G$ is what we are after: density contrast in the
absence of stellar contamination.  The second term amounts to a
small additional source of noise.  Since, as mentioned in \S
\ref{sec:likelihood}, the noise is completely unimportant on the
scales of interest, we neglect this term.  Therefore,
\be
\langle \Delta_i^G \Delta_j^G \rangle = (1-\alpha)^{-2}
\langle \Delta_i \Delta_j \rangle.
\ee
We have accordingly corrected all our $\C_B$ estimates
and their error bars upwards by $(1-\alpha)^{-2}\approx 1.29$.

By selection function we mean $d \bar G / dz$ where $\bar G$ is the mean
number of EDSGC galaxies per steradian.  The smooth curve in 
Fig.~\ref{fig:selection} was chosen
to fit the histogram, and is given by
\be
\frac{d\bar G}{dz}=4\times 10^5 \exp(-(z/0.06)^{3/2}) 
\left (\frac{z}{0.1}\right )^3. 
\ee 

Restricting ourselves to $b_J<19.4$ leaves around 200,000 galaxies.  
Although this is only $\sim 15\%$ of
the total number of galaxies in the EDSGC, the resulting shot noise
is still less than the fluctuation power, even at the smallest
scales we consider.

We binned the map into 5700 pixels with extent $0.5\deg$ in declination
and $0.5\deg$ in right ascension (RA).  The pixels are slightly
rectangular with varying solid angles; 
the RA widths correspond to angular distances 
ranging from $0.46\deg$ at $\delta = -23\deg$ to $0.37\deg$ at 
$\delta = -42\deg$.  This pixelization is fine enough so as
not to affect our interpretation of the large--scale fluctuations;
it causes a $\sim$4\% suppression of the fluctuation power at $l=80$. 
We have varied the pixelization scale to test this and find that
with $1\deg\times 1\deg$ pixels the estimated $\C_l$s change by less than 
half an error bar for $l < 80$.

\begin{figure*} [htb]
\centerline{\psfig{figure=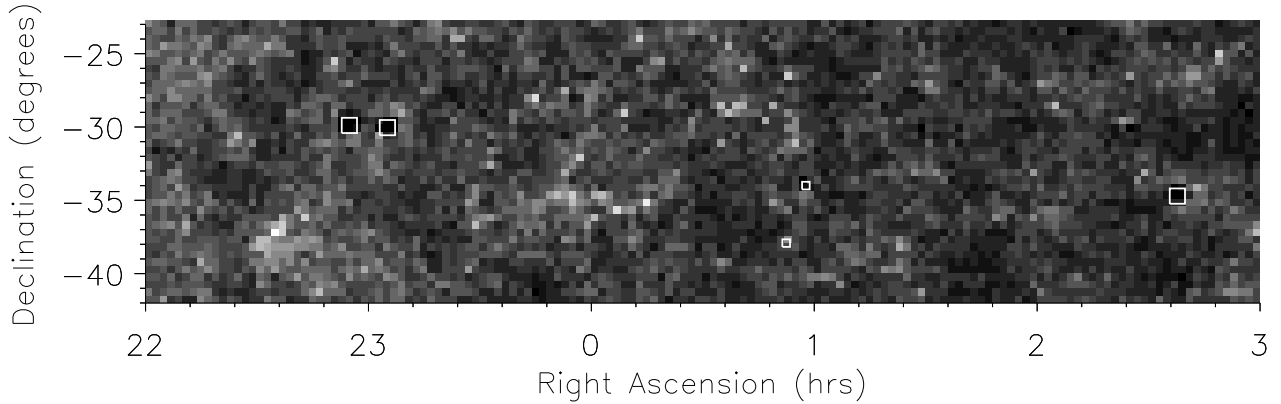,width=18cm}}
\caption{Map of the EDSGC that we used in our analysis ($23<{\rm
ra} <3$ hours and $-42<{\rm dec}<-23$ hours, $b_J<19.4$).  Five
of the largest masks are indicated with squares.  }
\label{fig:map}
\end{figure*}

We also took into account the ``drill holes'', locations in the
map which were obstructed (e.g. by bright stars). In the case of
$0.5\deg\times 0.5\deg$ pixelization about 75 pixels were
corrupted by drill holes. Those pixels were assigned large
diagonal values in the noise matrix (e.g., Bond, Jaffe \& Knox
1998), and thus had negligible weight in the subsequent analysis.
The $0.5\deg\times 0.5\deg$ pixelized map is shown in
Fig.~\ref{fig:map}.

\begin{figure}[htbp]
\begin{center}
\psfig{file=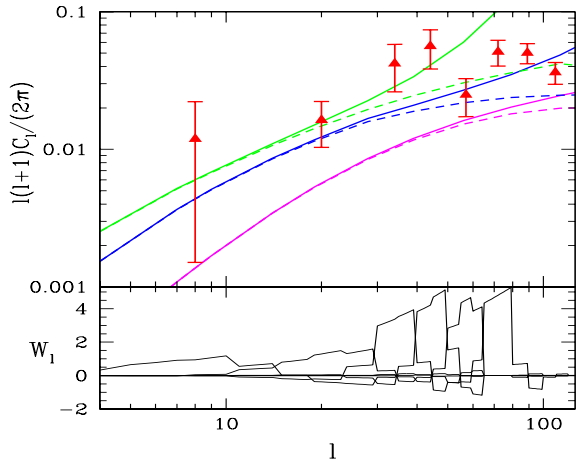 , height=2.625in, width=3.4in}
\caption{Angular power spectra estimated from the data and
predicted for various models.  From bottom to top at low $l$ are
the COBE and cluster consistent predictions for
$\Omega_m=0.15,0.3,1$ and $b=1$ (dashed lines show linear theory
predictions).  The lower panel shows the window functions for the
first six bands.  }
\label{fig:cl}
\end{center}
\end{figure}

In Fig.~\ref{fig:cl} we plot the estimated angular power spectrum
from the EDSGC data.  
Also shown in Fig.~\ref{fig:cl} are our predicted $\C_l$s.  
For each of these, we can calculate the expected values of $\C_B$
by summing over the window functions, shown in the bottom panel
for the six lowest $l$ bands.  The jaggedness results from our
practice of calculating the Fisher matrix not for every $l$,
but for fine bins of $l$ labeled by $b$. 
We then assume $F_{ll'} = F_{bb'}/(\delta l(b)\delta l(b'))$.

\begin{figure}[htbp]
\begin{center}
\psfig{file=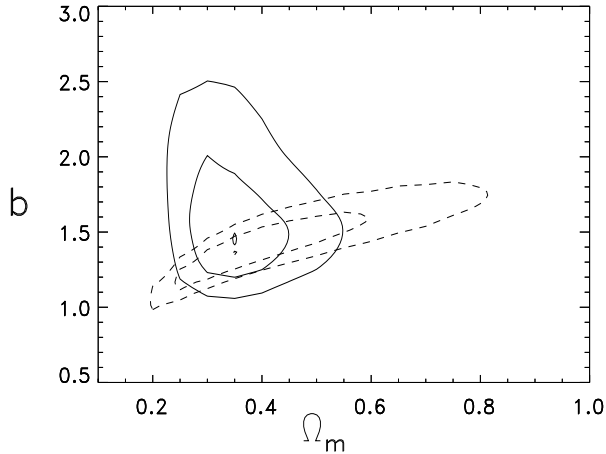 , height=2.625in, width=3.4in}
\caption{Contours of constant $\chi^2$ in the $\Omega_m$
vs. bias plane.  The dashed lines are for $\sigma_8$ chosen
to be at Viana and Liddle maximum-likelihood value.  The
solid line is the result of marginalizing over $\sigma_8$, with
the VL99 prior and a prior in $n$ of $1\pm 0.1$.
The contour levels show the minimum as well as 2.3 and 6.17
above the minimum, corresponding to 68\% and 95.4\%
confidence levels if the distribution were Gaussian.
}
\label{fig:contour}
\end{center}
\end{figure}

We apply equation (\ref{eqn:chisqEDSGC}) with the sum restricted
to the six $\C_B$ at lowest $l$.  First we keep $\sigma_8^c$ fixed to the
preferred value of 0.56 (VL99) resulting in a $\chi^2$ whose
contours are shown as the dashed lines in Fig.~\ref{fig:contour}.  
The minimum of this $\chi^2$ is 8.1 for $6-2=4$ degrees of freedom 
at $\Omega_m = 0.35$ and $b=1.3$, where $n=0.91$. This is an acceptable
$\chi^2$; the probability of a larger $\chi^2$ is 9\%. 
Moving towards higher $\Omega_m$ decreases the VL99 preferred value of
$\sigma_8$, and thus the preferred value of $b$ increases.  
Increasing $\Omega_m$ also changes the transfer function, requiring
a decrease in $n$ in order to agree with both 
{\it COBE}/DMR and cluster abundances.  This change in the shape of the angular
power spectrum leads to an increase in $\chi^2_{\rm EDSGC}$.  
Moving towards lower $\Omega_m$ generates a 
bluer tilt to the $C_l$ shape in two different ways.  It leads
to higher $n$ for consistency with {\it COBE}/DMR and cluster
abundances and it also increases
the importance of non--linear corrections.  These combined effects
lead to a rapidly increasing $\chi^2_{\rm EDSGC}$ for $\Omega_m < 0.2$.

The uncertainties on $\sigma_8^c$ from cluster abundances 
(as we interpret them) are
significantly larger than the EDSGC constraints on $b$ for fixed
$\sigma_8^c$.  If we
take them into account, we must include additional prior
information in order to obtain an interesting constraint on the
bias.  Since (at fixed $\Omega_m$) changing $\sigma_8^c$ changes
$n$, prior constraints on $n$ will help to constrain
$\sigma_8^c$.  Therefore we work with the total $\chi^2= \chi^2_{\rm EDSGC}+\chi^2_{VL} + \chi^2_n$.  From a combined
analysis of {\it Boomerang}-98, {\it Maxima}-I and {\it COBE}/DMR data,
Jaffe et al.\ (2000) find $n = 1 \pm 0.1$; hence we adopt
$\chi^2_n = (n-1)^2/0.1^2$.  We 
marginalize the likelihood, which is proportional to
$e^{-\chi^2/2}$, over $\sigma_8^c$.

\begin{figure}[htbp]
\begin{center}
\psfig{file=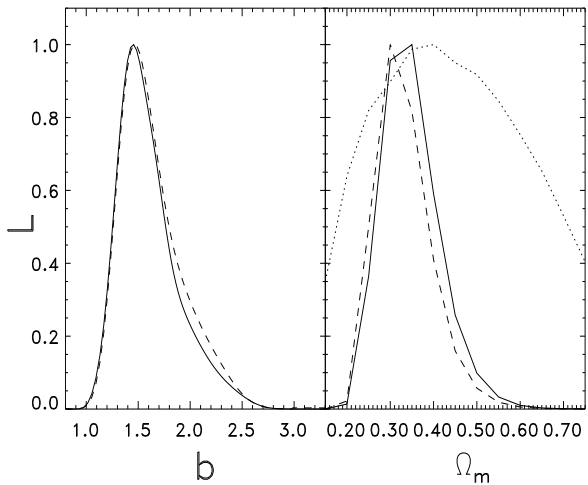 , height=2.625in, width=3.4in}  
\caption{Left panel: likelihood of $b$ marginalized over
$\sigma_8$ (with $n=1\pm 0.1$ prior) at $\Omega_m =0.35$ (solid
line), and additionally marginalized over $\Omega_m$ (dashed
line). Right panel: likelihood of $\Omega_m$ with no priors
(dotted line), $n$ prior (dashed line), $n$ and our VL priors (solid
line).  }
\label{fig:blike}
\end{center}
\end{figure}

Marginalizing over the amplitude constraint from cluster
abundances, we find $1.07 < b < 2.33$ at the best-fit value of
$\Omega_m = 0.35$, and $1.11 < b < 2.35$ after marginalizing over
$\Omega_m$ (both ranges 95\% confidence).  These constraints 
correspond to the solid and dashed contours respectively in
Fig.~\ref{fig:contour}.  Figure ~\ref{fig:blike} shows the
likelihood of bias, when marginalized either over $\sigma_8$
(solid line) or $\Omega_m$ (dashed line).
Marginalizing over the bias leads to weak constraints on
$\Omega_m$, unless one insists on allowing only small departures from
scale-invariance.  With the assumption that the primordial power
spectral index is $n=1\pm 0.1$, we find $0.2 < \Omega_m <
0.55$ at 95\% confidence.  Furthermore, it is interesting that
not only do ``concordance''--type models, with scale--independent
biases provide the best fits to the EDSGC data, but they also
provide acceptable fits. 

\section{Systematic Errors}\label{sec:sys}

In this section we discuss three sources of systematic error:
spatially--varying extinction by interstellar dust, deviation of
the survey mean from the ensemble mean and deviation from
Gaussianity.  Above we have assumed their impact on the data to
be negligible.  In the following we use maps with three different
pixelizations: BIGPIX (1.5 deg $\times$ 1.5 deg pixels, a total
of $N=650$ of them), MEDPIX (1.0 deg $\times$ 1.0 deg, $N=1425$)
and FINEPIX (0.5 deg $\times$ 0.5 deg, $N=5700$). Note that
FINEPIX was ultimately used to obtain the cosmological parameter
constraints.  Coarser pixelizations, however, are easier to work
with due to a much smaller number of pixels (in particular,
$N\times N$ matrices have to be repeatedly inverted in the
quadratic estimator).  

\subsection{Interstellar Dust}

The first possible source of systematic error, interstellar dust, we can
dispense with quickly due to the work of Nichol \& Collins (1993) and, more
recently, Efstathiou \& Moody (2000). The former investigated the effects of
interstellar dust (using HI and IRAS maps as tracers of the dust) on the
observed angular correlation function of EDSGC galaxies (see Collins et
al. 1992) and found no significant effect on the angular correlations of these
galaxies to $b_j=19.5$. We note that Nichol \& Collins (1993) also
investigated plate--to--plate photometric errors and concluded they were also
unlikely to severely effect the angular correlations of EDSGC
galaxies. Efstathiou \& Moody (2000) used the latest dust maps from Schlegel,
Finkbeiner \& Davis (1998) to make extinction corrections to the APM catalog
and found that for galactic latitudes of $|b|>20^\circ$, the corrections have
no significant impact on the angular power spectrum. Since all the EDSGC
survey area resides at galactic latitudes of $|b|>20^\circ$ and has been
thoroughly checked for extinction--induced correlations, we conclude that
spatially--varying dust extinction has not significantly affected our
power--spectrum determinations either.

\subsection{Integral Constraint}

We are interested in the statistical properties of deviations
from the mean surface density of galaxies.  This effort is
complicated by our uncertain knowledge of the mean.  Our best
estimate of the ensemble mean is the survey mean.  But assuming
that the survey mean is equal to the ensemble mean leads to
artificially suppressed estimates of the fluctuation power on the
largest scales of the survey.  This assumption is often referred
to as ``neglecting the integral constraint'' (for discussions,
see, e.g., Peacock \& Nicholson (1991); Collins, Nichol \&
Lumsden (1992)).

Let $\bar G$ be the ensemble average number of galaxies in a
pixel.  Let us denote the survey average as
\be
\tilde G = {1\over n_{\rm pix}} \sum_i G_i.
\ee
Since we do not know the ensemble average, in practice we use
the survey average to create the contrast map:
\be
\label{eqn:contrast}
\tilde \Delta_i = {G_i - \tilde G \over \tilde G} = {1 \over 1+\epsilon}
\left(\Delta_i -\epsilon\right)
\ee
where
\be
\Delta_i \equiv {G_i - \bar G \over \bar G} 
\ee
is the contrast map made with the ensemble average and 
\be
\epsilon \equiv {\tilde G - \bar G \over \tilde G}
\ee
is the fractional difference between the two averages
(for simplicity of notation we are assuming equal--area pixels).

Our likelihood function should not have the covariance matrix
for $\Delta_i$, but instead for $\tilde \Delta_i$.  These
are related by
\be
\label{eqn:corr_terms}
\langle \tilde \Delta_i \tilde \Delta_j \rangle =
\langle \Delta_i \Delta_j \rangle 
-\langle \epsilon\left(\Delta_i + \Delta_j\right)\rangle +
\langle \epsilon^2 \rangle
\ee
plus higher order terms\footnote{An exact
expression to all orders is given by equation (20) of
Gazta\~{n}aga \& Hui (1999).}.  The extra terms of 
the above equation are easily calculated
with the following expressions:

\bea
\langle \epsilon \Delta_i \rangle 
&=& {1\over N_{\rm pix}} \sum_j \langle \Delta_i \Delta_j \rangle \nonumber \\
\langle \epsilon^2 \rangle 
&=& {1\over N_{\rm pix}^2} \sum_{ij} \langle \Delta_i \Delta_j \rangle 
\eea

Each correction term typically contributes 10-20\% to the
corresponding terms of the covariance matrix (they do not cancel,
since there are {\it two} linear correction terms; see
equation (\ref{eqn:corr_terms})). The main contribution comes from the
lowest multipoles, corresponding to largest angles $\theta$.
Indeed, the correction terms come almost entirely from our lowest
multipole bin.  Dropping this bin (or using a $\Lambda$CDM $C_l$) reduces
the correction terms to 2\% or less.

\begin{figure}[htbp]
\begin{center}
\psfig{file=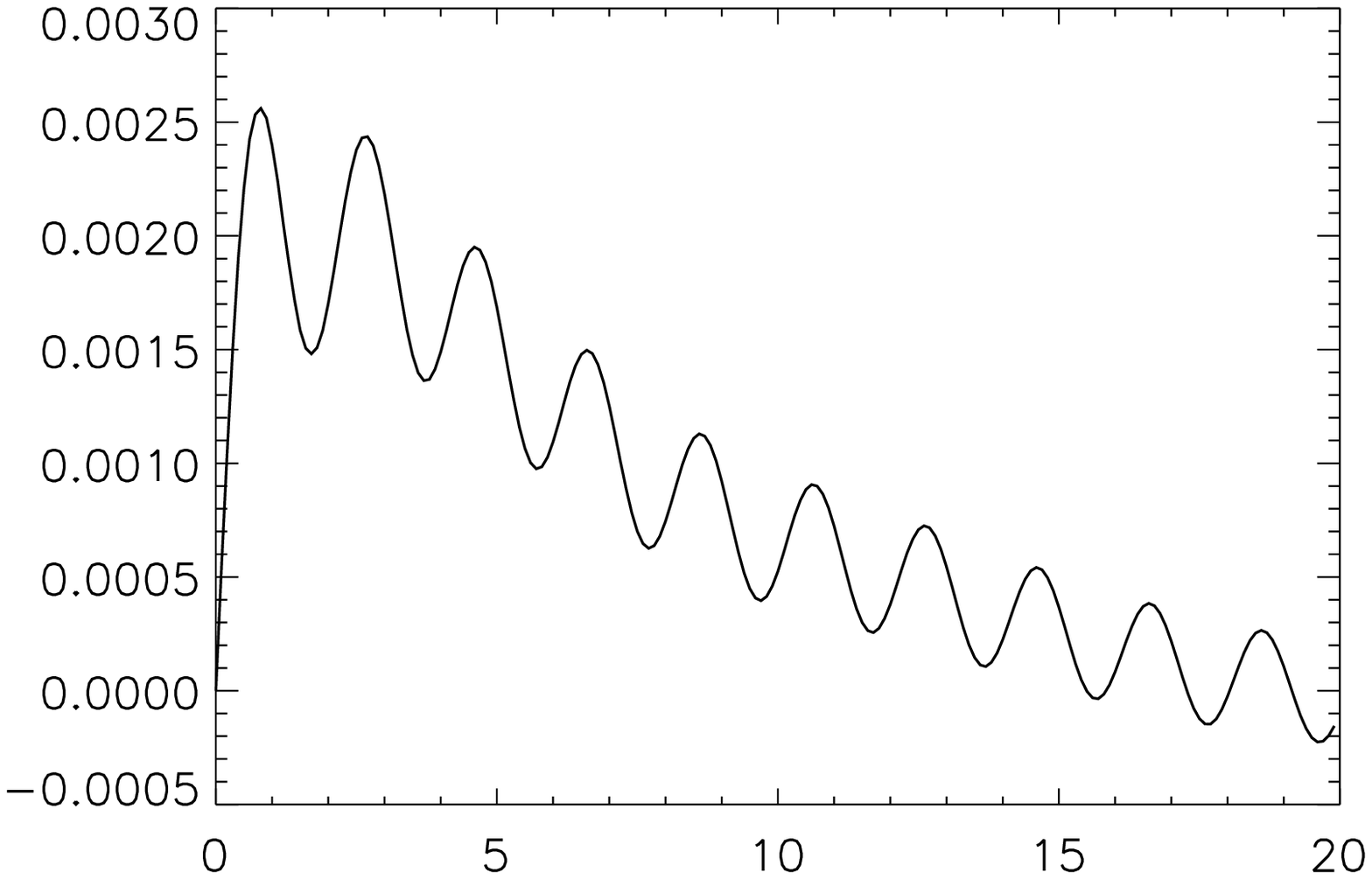, height=2.625in, width=3.4in}
       \put(-135,5){$\theta$ (degrees)} 
	\put( -250, 37){\rotatebox{90}{$2\pi \theta S(\theta)/
                           \Omega/S(0)$ (degrees$^{-1})$}}
\caption{The area under the curve is approximately equal to
the integral constraint correction terms of equation 
(\ref{eqn:corr_terms}) in  units of $S(0)$;
see equation (\ref{eqn:anaintcorr}).   The assumed model is
$\Omega_M=0.3$ with VL99 and {\it COBE}/DMR normalization.
(The oscillations are due to the fact that only the 
contributions from multipole moments at $l < 180$ were included.)}
\label{fig:s_theta}
\end{center}
\end{figure}

The amplitude of 
the correction terms can be understood from the weakness of the
signal correlations on scales approaching the smaller survey dimension 
of $19^\circ$.  In that case, we can write:
\be
\label{eqn:anaintcorr}
{1\over N_{\rm pix}} \sum_j \langle \Delta_i \Delta_j \rangle
\approx 2\pi \int S(\theta) \theta d\theta /\Omega.
\ee
where $\Omega$ is the area of the survey, and $S(\theta)$
is the signal covariance, given by the RHS of equation 
(\ref{eqn:signal_matrix})
(we have neglected pixel noise).  We plot the integrand
in Fig.~\ref{fig:s_theta} in units of $S(0)$.  

\begin{figure}[htbp]
\begin{center}
\psfig{file=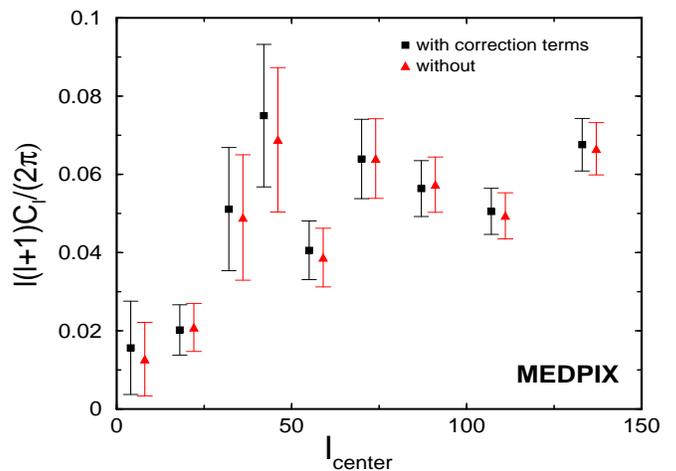, height=2.5in, width=3.4in}
\caption{The $\C_l$ determined with and without the
integral constraint correction. Shown is the MEDPIX case, and
abscissae of points were slightly offset for easier viewing.}
\label{witheps_noeps}
\end{center}
\end{figure}

Fortunately, even though the correction terms are not
entirely negligible, their inclusion makes the estimated $\C_l$ change
very little. This is shown in Fig.~\ref{witheps_noeps}.  The most
significant change is a $\sim 20$\% broadening of the error bar of the
lowest multipole.  Including this effect has a negligible consequence
on our cosmological parameter constraints.

\subsection{Gaussianity}

\begin{figure}[htbp]
\begin{center}
\psfig{file=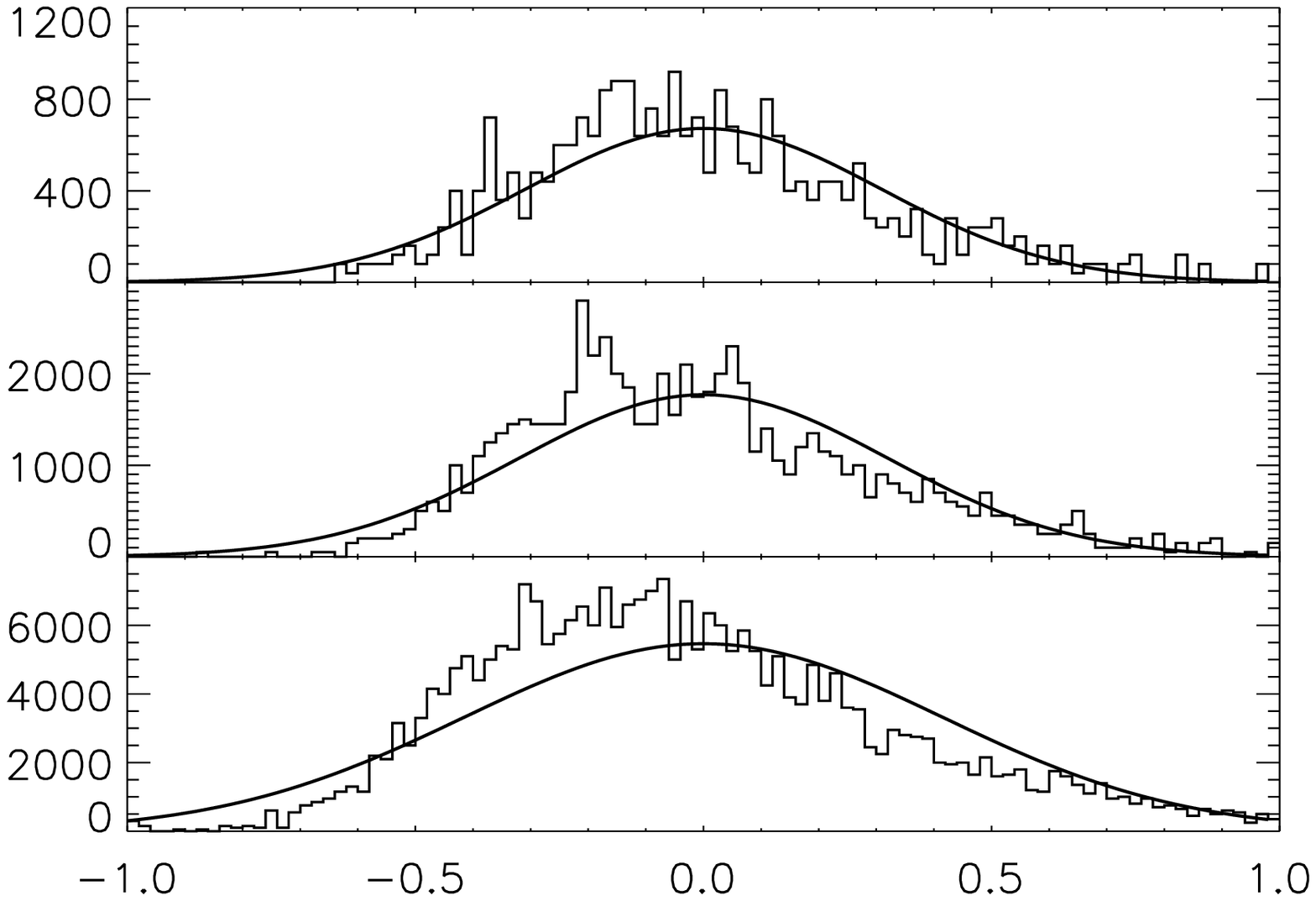, height=2.625in, width=3.4in}
       \put(-130, -10){$\Delta$}
        \put( -265, 84){\rotatebox{90}{$dN/d\Delta$}}
\caption{Histograms of the data, overplotted with Gaussians
centered at zero with variances equal to the pixel variances, for
maps made with three different pixel sizes.  From top to bottom 
they are: BIGPIX, MEDPIX, FINEPIX.}
\label{3panel.hist}
\end{center}
\end{figure}

On large enough scales, we expect the maps to be Gaussian--distributed.
Figure ~\ref{3panel.hist} shows histograms of the data for the three 
pixelizations we examined.  The histograms are overplotted with
the Gaussians with zero mean and variance equal to the pixel variance.
One can see the improved consistency with Gaussianity as the
pixel size increases.  

We applied a Kolmogorov--Smirnov test (e.g., Press et al.\ 1992)
to check for
consistency of the above histograms with their corresponding
zero--mean Gaussians.  We find probabilities that these
Gaussians are the parent distributions of $<10^{-10}$, 0.001, and
4.5\% for FINEPIX, MEDPIX and BIGPIX respectively, indicating
that Gaussianity is a better approximation on large scales than
it is on small scales, as expected.  We also determined the
skewness of the maps in units of the variance to the 1.5 power,
and find the same trend of decreasing non--Gaussianity with scale:
1.21, 0.85, 0.79.

The trend with increasing angular scale and the weakness
of the $\sim 2 \sigma$ discrepancy for the BIGPIX map are reassuring
for our analysis that considered only moments $l < 80$.  Note that a 
spherical harmonic with $l = 80$ has 3 BIGPIX pixels in 
a wavelength.  However, a normalized skewness near unity is
worrisome---and this skewness is not decreasing rapidly with
increasing angular scale.  We discuss possible ways of dealing
with this non--Gaussianity in the next section.

\section{Discussion}\label{sec:discussion}

We reduced our sensitivity to the non--Gaussianity of the data by restricting
our cosmological parameter analysis to $l < 80$.  However, the map may still
be significantly non--Gaussian even on these large scales.  Future analyses of
more powerful data sets that result in smaller statistical errors will have to
quantify the effects of the Gaussianity assumption, which we have not done
here.

The non--Gaussianity may force us towards a Monte--Carlo approach.  An
analysis procedure similar to the one utilized here may have to be repeated
many times on simulated data---where the simulations include the non--linear
evolution that presumably is the source of the Gaussianity.  The distribution
of the recovered parameters can then be used to correct biases and
characterize uncertainties.

Monte--Carlo approaches may be necessary for other reasons as well.  Recently
Szapudi et al.\ (2000) have tested a quadratic estimator for $C_l$ with a
simpler (sub--optimal) weighting scheme that only requires on the order of
$N^2$ operations (or $N\sqrt{N}$ operations using the new algorithms of Moore
et al. 2001) instead of $N^3$.  A drawback is that evaluation of analytic
expressions for the uncertainties requires on the order of $N^4$ operations.
Fortunately, the estimation of $C_l$ is rapid enough to permit a Monte--Carlo
determination of the uncertainties in a reasonable amount of time.

Note though that Bayesian approaches may still be viable, if it can be shown
that non--Gaussian analytic expressions for the likelihood provide an adequate
description of the statistical properties of the data.  See Magueijo, Hobson
\& Lasenby (2000) and Contaldi et al.\ (2000).

To get our constraints on cosmological parameters we fixed the Hubble constant
at 70 km/sec/Mpc, or $h=0.7$.  We now explain how our bias results and
$\Omega_m$ results scale for different values of the Hubble constant.

The transfer function depends on the size of the horizon at
matter--radiation equality $\lambda_{\rm EQ}$ which is
proportional to $1/(\Omega_m h^2)$, or, in convenient distance
units of $h^{-1}{\rm Mpc}$, $1/(\Omega_m h)$.  The latter quantity
is the relevant one since all distances come from redshifts and
the application of Hubble's law (in this case the redshifts taken
for our selection function) with the result that distances are
only known in units of $h^{-1}$Mpc.  Thus there is a degeneracy
between models with the same value of $\Omega_m h$ and different
values of $h$.

This degeneracy is broken by the $\Omega_m$ dependence of the
COBE-normalization of $\delta_H$ and the cluster normalization of
$\sigma_8$.  Increasing $h$ at fixed values of $\Omega_m h$ means
$\Omega_m$ decreases, raising both $\delta_H$ and $\sigma_8$.
Ignoring non--linear effects, this can be mimicked by an increase
in the bias and only a very slight reddening of the tilt (since
$\delta_H$ has risen only slightly more than $\sigma_8$ and there
is a long baseline to exploit).

The end result is that our constraints on $b$ are actually constraints on 
$b(h/0.7)^{-0.5}$, and our constraints on $\Omega_m$ (at least when 
marginalized over bias) are actually constraints on $\Omega_m (h/0.7)$.  

\section{Conclusions}\label{sec:conclusions}

We have presented a general formalism to analyze galaxy surveys
without redshift information.  We pixelize the galaxy counts on the
sky, and then, using the quadratic estimator algorithm, extract
the angular power spectrum -- a procedure already in use in CMB
data analysis. Just like in the CMB case, one effectively
converts complex information contained in the experiment (in this
case, locations of several hundred thousand galaxies) into a
handful of numbers -- the angular power spectrum. One can then
use the angular power spectrum for all subsequent analyses. 

We apply this method to the EDSGC survey. We compute the angular
power spectrum of EDSGC, and combine it with {\it COBE}/DMR and
cluster constraints to obtain constraints on cosmological
parameters. Assuming flat $\Lambda$CDM models with constant bias
between galaxies and dark matter, we get $1.11 < b < 2.35$ and
$0.2 < \Omega_m < 0.55$ at 95\% confidence. 

One advantage of our formalism is that it does {\it not} require
galaxy redshifts, but only their positions in the sky. This
should make it useful for surveys with very large number of
galaxies, only a fraction of which will have redshift
information. For example, the ongoing SDSS is expected to collect
about one million galaxies with redshift information, but also a
staggering one hundred million galaxies with photometric
information only. Using the techniques presented in this paper,
one will be able to convert that information into the angular
power spectrum, which can then be used for various further
analyses.  

\acknowledgments 

We thank Scott Dodelson, Daniel Eisenstein, Roman Scoccimarro and
Idit Zehavi for useful conversations and J. Borrill for use of
the MADCAP software package.  DH is supported by the DOE. LK is
supported by the DOE, NASA grant NAG5-7986 and NSF grant
OPP-8920223.  RN thanks NASA LTSA grant NAG5-6548.

\appendix

\section{Limber's Equation}\label{app-limber}

In order to derive the equation giving $C_l$ as a function of
$P(k)$ we must understand the dependence of the data on the 3D
matter density contrast $\delta \equiv \delta \rho/\rho$ as a
function of time and space.  First, we relate the number of
galaxies per unit solid angle $G$ observed from location $\r$
in a beam with $FWHM=\sqrt{8\ln{2}}\sigma$ centered on the direction
$\gam$ to the comoving number density of
detectable galaxies $g$, via

\be
G(\r,\gam) = \int d^3r' { e^{-|\hat x-\gam|^2/2\sigma^2}
\over 2\pi \sigma^2} g(\r',\tau_0 - x), 
\ee

\noindent where $\x \equiv \r'-\r$, $x$ is the magnitude of $\x$
and $\tau_0$ is the conformal distance to the horizon today.  To
relate $g$ to $\delta$ we simply assume that the galaxies are a
biased tracer of the mass, so that $g = \bar g(1+b\delta)$.
Therefore:
\bea
\Delta(\r,\gam) 
&\equiv& {G-\bar G \over \bar G} \nonumber \\ 
&&\hspace{-2cm} = {1\over \bar G} \int d^3r' { e^{-|\hat x-\gam|^2/2\sigma^2}
\over 2\pi \sigma^2} \bar g(x)b(x)\delta(\r',\tau_0-|x|),
\eea
where we have allowed for a time-dependent (and therefore $x$-dependent)
bias.  If we further assume that the density contrast grows uniformly
with time, with growth factor $D(x)$, then we can write:

\be
\label{eqn:signal}
\Delta(\r,\gam) = {1 \over \bar G} \int d^3r' { e^{-|\hat x-\gam|^2/2\sigma^2}
\over 2\pi \sigma^2} \bar g(x)b(x)\delta(\r',\tau_0)D(x). 
\ee

Calculating 
$w(\theta_{12}) = \langle \Delta(\r,\gam_1)\Delta(\r,\gam_2) \rangle$ 
and then taking its Legendre transform yields (after a fair amount of algebra):
\be
\label{eqn:Limber}
C_l = {2\over \pi} \int k^2 dk P(k) f_l(k)^2
\ee
where
\be
f_l(k) \equiv {1 \over \bar G}\int {dx\over F(x)} j_l(kx)x^2\bar g(x) D(x) b(x)
\label{eqn:f_l}
\ee
and $F(x)$ enters the metric via

\be
ds^2 = a^2\left[d\tau^2 - dx^2/F(x) + x^2 d\theta^2 + x^2 \sin^2\theta d\phi^2
\right].
\ee

\noindent For zero mean curvature $F(x)=1$; expressions valid for 
general values of the curvature are given by Peebles (1980, equation (50.16)).

Note that 
\bea
\bar G & = &\int {r^2 \over F(r)} \bar g(r) dr \nonumber\\
&=& \int dz {d\bar G \over dz}
\eea
and therefore
\be
{r^2 \over F(r)} \bar g(r) dr/dz = 
{ d\bar G \over  dz}.
\ee

One can use equations (\ref{eqn:Limber}) and (\ref{eqn:f_l}) to calculate
the expected value of $C_l$ for any theory.  The only information
one needs from the survey to do this is $\bar g(r)$ or ${d\bar
G/dz}$.  The latter is preferable, and what we use in our
application, because it is directly observable as long as
redshifts in some region are available.

\end{document}